\documentstyle[aps,pra,epsf]{revtex}

\begin{document}

%%%%%%%%%%%%%%%%%%%%%%%%%%%%%%%%%%%%%%%%%%%%%%%%%%%%%%%%%%%%%%%%%
% inserted for two column style
\twocolumn[\hsize\textwidth\columnwidth\hsize\csname @twocolumnfalse\endcsname

\draft
\widetext 
\title{A proper ballistic calculation of tunneling conductance for real junctions}
\author{P. M. Levy$^{a}$, K. Wang$^{a}$, P. H. Dederichs$^{b}$, C.
Heide$^{a}$, S. Zhang$^{c}$, L. Szunyogh$^{d}$, and P. Weinberger$^{e}$}
\address{$^{a}$Department of Physics, New York University, 4 Washington Place, New\\
York, NY 10003\\
$^{b}$Forschungszentrum J\"{u}lich, Institut f\"{u}r Festk\"{o}perforschung,\\
D-52425, J\"{u}lich, Germany\\
$^{c}$Department of Physics and Astronomy, University of Missouri, Columbia,\\
MO 65211\\
$^{d}$Department of Theoretical Physics, Technical University Budapest,\\
H-1521 Budapest, Hungary\\
$^{e}$Institut f\"{u}r Technische Electrochemie, Technische Universit\"{a}t\\
Wien, A-1060 Wien, Austria}
\date{\today}
\maketitle

\begin{abstract}
Employing an {\it ab initio} Screened Korringa-Kohn-Rostoker (SKKR) band
structure method for a metal-vacuum-metal junction, we find that the tunnel
conductance is different when it is calculated across the barrier and far from
it. We attribute this difference to an artefact of the ballistic approach
which overestimates the role of specular reflections, and its inability to
pick up contributions from localized interface states. To reconcile the
ballistic approach with experiment, we propose that the tunnel conductance
should be calculated {\it as if} it is measured directly across the barrier.
In this case the predicted tunneling magnetoresistance is larger.
\end{abstract}

\pacs{PACS numbers: 73.40.Gk, 75.70.-i, 73.20.At, 75.70.Pa}
\phantom{.}
] \narrowtext
%%%%%%%%%%%%%%%%%%%%%%%%%%%%%%%%%%%%%%%%%%%%%%%%%%%%%%%%70%%%%%%%
Measurements of the resistance and magnetoresistance (MR) of magnetic planar
tunnel junctions are usually made by passing a current with voltage probes
far removed from the interfaces between the electrodes and insulating
barrier. Therefore it makes sense to calculate these transport properties as
the transmission probability from propagating eigenstates in one electrode
to those of the other. To maintain a steady state current the electrodes are
connected to reservoirs, and it is understood that propagating eigenstates
are determined far from the electrode/barrier interfaces. While this
procedure is reasonable, a consensus based on data and intuitive grounds has
evolved that the current in tunnel junctions is controlled by the electronic
structure at the interfaces, e.g., the local density of states (LDOS)~\cite
{moodera}. This conflicts with the above description which uses the
electronic structure in the electrodes far from the interfaces, which is
different from that at the interface. In this letter we point out that
calculations made with the Landauer-B\"{u}ttiker or other formalism for
purely ballistic transport over the {\it whole} junction cannot be directly
compared to data on real junctions for at least two reasons. The ballistic
conductance overestimates the role of specular reflections from the barrier
as seen by the electrodes, and, if they are present, overlooks contributions
from states localized near the interfaces that are coupled to itinerant
states in the electrodes by diffusive and relaxation processes in real
junctions. The first causes the ballistic conductance to decrease much more
rapidly than it would with diffusive electrodes; the second provides
additional conduction channels. Under these conditions the electronic
structure at the electrode/barrier interfaces controls the tunneling
current. As we will show, to model the conductance of real planar tunnel
junctions, i.e., with diffusive electrodes, it should be calculated {\it as
if} it is measured directly across the barrier. In other words, to reconcile
the ballistic approach with experiment one has to change the boundary
conditions on the transport calculation.

%%%%%%%%%%%%%%%%%%%%%%%%%%%%%%%%%%%%%%%%%%%%%%%%%%%%%%%%70%%%%%%% 
\begin{figure}[tbp] 
\begin{center} 
\leavevmode
{\hbox 
{\epsfxsize=3.3in
\epsffile{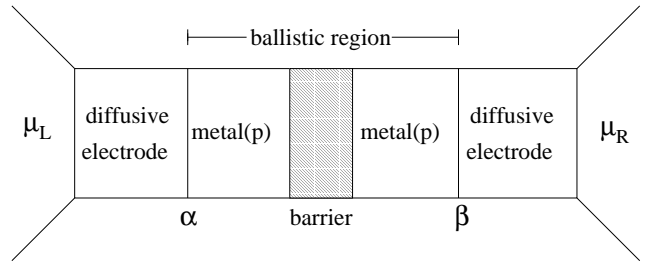}}}
\vspace{+1ex}
\caption{Scheme for conductance calculations across different ballistically 
conducting sections of a transition metal/vacuum/transition metal junction. 
At the points $\protect\alpha$ and $\protect\beta$ the ballistic region is 
joined with diffusive electrodes thus forming an open system with reservoirs 
on each side.}  \end{center} 
\end{figure} 

We have calculated the conductance of transition-metal/vacuum tunnel
junctions by using the Caroli formalism\cite{caroli}. While real junctions
have insulating barriers, we have taken a vacuum as it is the simplest
insulator for which we can do an {\it ab-initio} calculation. Normally, one
sets the chemical potentials far from the barrier, i.e., deep in the leads.
In such an approach  one does not specify the chemical potentials at
intermediate planes across which one chooses to evaluate the current, and 
one keeps the ballistic information about the propagators throughout the
entire junction; not just between these planes. In our approach we set the
chemical potentials at the planes we are looking at $\mu _{\alpha (\beta
)}\simeq \mu _{L(R)}$, and so far as transport is concerned we are able to
isolate the region between $\alpha $ and $\beta $ from the remainder of the
junction; see Fig.1. Thus, as we see below, one is able to talk about the
conductance of a part of a full junction. 

In the linear response region, the tunneling conductance is~\cite{kuising} 
\begin{equation}
G=\frac{2\pi e^{2}}{\hbar }\mbox{Tr}\left\{ \rho ^{\alpha }(\epsilon
_{F})[t^{\dagger }(\epsilon _{F})]^{\alpha \beta }\rho ^{\beta }(\epsilon
_{F})t(\epsilon _{F})^{\beta \alpha }\right\} ,  \label{4}
\end{equation}
where $t^{\alpha \beta }$ is the t-matrix between $\alpha $ and $\beta $;
the trace Tr is over all (site and angular momentum, or energy level)
indices; and the density of states (DOS) at the Fermi level $\rho ^{\alpha }$
and $\rho ^{\beta }$ used in this formalism are those at a surface created
by cutting the junction so that the two parts thereby created are isolated
from one another\cite{pollmann}.

In order to obtain the t-matrix, we first determine the propagator $G$ from
a full junction calculation and then backward derive the t-matrix from
Dyson's equation $G=g+gVG=g+gtg$, where $g$ is the propagator in the absence
of the perturbation $V$. In our case, $V$ joins diffusive electrodes with a
ballistically conducting system consisting of $p$ magnetic monolayers of a
transition metal, the electrodes, on each side of a vacuum barrier of six
monolayers as shown in Fig.1. We adopted an empty lattice structure for the
vacuum layers which has the same properties as the metallic electrodes
except that it contains no atoms. By definition $g^{\alpha \beta }$ is zero
and, due to the fact that $V$ takes the form of a nearest neighbor
interaction (in a principal layer description in which we use 2 atomic
layers we have indeed nearest and next nearest neighbor interactions \cite
{Laci}), we find\cite{kuising}

\begin{equation}
G^{\alpha \beta }=g^{\alpha \alpha }t^{\alpha \beta }g^{\beta \beta }.
\label{5}
\end{equation}
Upon inversion this yields the t-matrix, which differs for different $\alpha 
$ and $\beta $. By inserting this result in Eq.~(\ref{4}) to calculate the
conductance we treat the transport between $\alpha $ and $\beta $
ballistically inasmuch as we explicitly use the propagator $G^{\alpha \beta
} $, while the transport in the regions of the electrodes to the left of $%
\alpha $ and the right of $\beta $ are treated diffusively, because we do
not keep track of the momentum there. Rather their effect on conduction is
taken into account in a ``mean field-like'' manner by a self energy term in
the propagator \cite{datta}. In the limit of large $p$ the conductance
converges to a system that is independent of $p$, i.e.\ the full ballistic
junction.

The conductance for bcc(100) Fe/vacuum/Fe, and for fcc(100) Co/vacuum/Co
tunnel junctions has been calculated from band structures obtained from the
spin-polarized scalar-relativistic Screened Korringa-Kohn-Rostoker (SKKR)
method\cite{Laci}, and the atomic sphere approximation (ASA) is used. 
Here we present the results for Fe; they are further coroborated by those on
Co. The lattice parameter for Fe is 5.27 a.u. (atomic units). Two
atomic layers are included in each screened principal layer, and the
screening potential is set to $2$ Ry inside each atomic cell. The
Gunnarsson-Lundqvist\cite{Gunnarsson} exchange-correlation potential is
used, and energy integration is performed by means of Gaussian quadrature
with 16 points on a semi-circle in the upper half complex energy plane. For
self-consistent calculations of the bulk metal, the free metal surface and
the metal-vacuum-metal interface potentials, 45 ${\bf k}_{\parallel }$
points are used in the irreducible wedge of two dimensional Brillouin Zone
(2DBZ), which enables the Fermi level to be converged up to $10^{-7}$ Ry.
For more details on this method see Ref.~\cite{Laci}. We used a small
imaginary part of the energy of $\varepsilon =0.5mRyd$ in the propagators in
order to converge our results in a reasonable time.

%%%%%%%%%%%%%%%%%%%%%%%%%%%%%%%%%%%%%%%%%%%%%%%%%%%%%%%%70%%%%%%% 
\begin{figure}[tbp] 
\begin{center} 
\leavevmode
{\hbox 
{\epsfxsize=3.0in
\epsffile{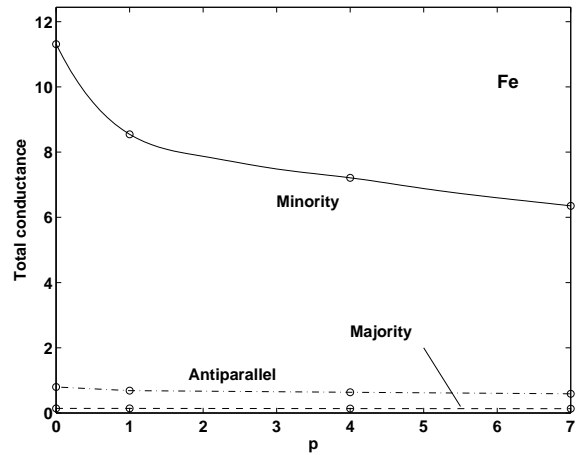}}
}
\vspace{+1ex}
\caption{Total conductance calculated in units of $e^{2}/\hbar$ $\times10^{-4}$
across different sections of a Fe(p)/vac(6)/Fe(p) tunnel junction with
$\protect\varepsilon=0.5mRyd$.}  
\end{center}  
\end{figure} 

In Fig.2 we show the ballistic conductance for Fe(p)/vac(6)/Fe(p) tunnel
junctions calculated either across the barrier, i.e.\ $p=0$, or up to $p=7$
Fe layers from the barrier. The conductance in all channels, majority, minority
and antiparallel, decrease as we are increasing the number of monolayers of the
electrodes we include in the ballistic calculation; this is particularly
pronounced in the minority channel. Most of the decay in the ballistic
conductance is between $p=0$ and $1$, because this is the region where the
electronic structure is changing most. It stands to reason that the usual
approach, in which the entire junction is treated ballistically, produces a
lower conductance than the one we would calculate just across the barrier. The
full ballistic approach includes the additional specular reflections near the
barrier that comes from the bending of the band bottoms which represents the
charge that has leaked out of the metallic electrodes into the vacuum barrier.
For example, we find 0.7 of a total of 3 minority electrons leak
out of the surface layer of the Fe electrode; for the majority band only 0.1
of the 5 electrons leak out. In Fig.2 one sees that the effect of the leakage
is far more pronounced in the minority band than the majority, and this
explains why the conductance drops much more in the minority band than the
majority when we take into account the reflections due to the bending of the
band bottoms.

In real junctions transport is diffusive, particularly in the electrodes, so
that the total resistance is the sum of resistances of each part; in this
case one should limit a ballistic calculation to just the barrier. As the
resistance of the barrier in tunnel junctions is about $10^{5}$ times
greater than that of the electrodes, even for junctions with resistances as
low as $50\Omega \mu m^{2}$, it is reasonable to say that the barrier
determines the resistance ($p=0)$ and the electrodes give negligible
contributions, i.e., in real planar junctions the conductance does not vary
much as one goes away from the barrier. The decay of the ballistic
conductance for increasing $p$ is thus an artefact of having considered the
transport in the electrodes ballistically.

We note that the tunneling magnetoresistance MR is larger when it is
measured across the barrier. While we obtain MR ratios $(G_P-G_{AP})/G_{P}$
of 82\% and 38\% for the Fe and Co junctions when we use the conductances
calculated at $p\rightarrow \infty $, these ratios increase to 86\% and 65\%
when measured directly across the barrier, i.e., at the interfaces $p=0$. 

In addition, as we will show now for the case of the Fe/vacuum junction,
diffuse scattering and relaxation processes may couple localized states at
the interfaces to propagating states in the electrodes. As in Fe these
localized states are dominant in the minority channel close to the Fermi
energy, they give an additional rise to the MR. The complete eigenvalue
spectrum of a semi-infinite solid contains the continuum of bulk states as
well as additional states localized at the surface~\cite{zangwill}. In
ferromagnetic metals such as Fe(100) the existence of localized states at the
surface has been known for some time~\cite{stroscio}; more recently surface
states were also observed in Gd(0001)~\cite{bode}. If the energy of the
surface states lie in the gaps between the bands of the bulk states they form
true localized states; otherwise they are resonant states - admixtures of
itinerant and localized states. Tunneling is certainly affected by these {\it
resonant} states at the interfaces of the junction; see for example Fig.2b in
Ref.~\cite{butler} where these resonances appear along the $\Gamma -M$
direction away from the zone center. The true localized states, on the
contrary, are orthogonal to resonant and itinerant states; therefore ballistic
transport will be unaffected by them. However, it has been shown that the
conductance through the localized states at Fe/vacuum and Gd/vacuum interfaces
can be substantial because these states have orbits that point out from the
surface into the barrier\cite{stroscio,bode}. Therefore it is the diffusive
nature of the transport and the ambient relaxation that allows the localized
states to contribute to conduction.

At $T=0K$ electrons can scatter elastically from the itinerant states only
to localized states at the Fermi energy and {\it \ vice versa}; this could
come from impurities and roughness of the interface. The scattering produced
by impurities, e.g., in Fe, was calculated by Mertig to be about $1\mu
\Omega cm/atomic\%$\cite{mertig2}; therefore, for $1atomic\%$ impurities, we
can roughly estimate the elastic scattering produced at the interfaces is
equivalent to the scattering rate of the order of $1/\tau _{imp}=10^{14}\sec
^{-1}$\cite{ashcroft}.
For the tunnel junctions studied to date with
resistances in the range of $10^{3}-10^{8}\Omega \mu m^{2}$, as well as for
the Fe/vac/Fe junction we will discuss, the tunneling rate is in the range $%
10^{6}-10^{10}\sec ^{-1}$, so that the diffuse scattering in sufficient to
have localized states contribute to participate in conduction. At finite
temperatures relaxation processes, such electron-electron, electron-phonon,
and electron-magnon interactions, can couple the localized to itinerant
states; our rough estimates tell us that the relaxation is faster than the
tunneling rate of $10^{6}\sec ^{-1}$when $T>0.7K$, while one needs $T\sim 20K
$ for the localized states to participate in the conduction when the
tunneling rate is $10^{10}\sec ^{-1}$. We conclude that while tunneling
electrons create holes in the localized interface states, they recombine
almost instantaneously due to scattering by phonons, magnons, other
electrons, or interfacial disorder so that for the calculation of the current
one can always assume a Fermi distribution even for the localized states. On
the contrary for currents perpendicular to the plane of the layers (CPP) in
metallic multilayered structures the rate at which electrons traverse a layer
is determined by the Fermi velocity, which is the of the order of $10^{16}\sec
^{-1}$. As the relaxation mechanisms are much slower, it is reasonable to
calculate the conductance in metallic multilayers by neglecting relaxation
to localized states, even though they appear at interfaces in much the same
way as surface states~\cite{mertig}.

At the Fe(100)/vac interface surface states exist in the minority channel at
the Fermi level for $k_{\parallel }\neq 0$; for $k_{\parallel }\sim 0$
localized states exist above $\epsilon _{F}$~\cite{stroscio,julich};
therefore those with $k_{\parallel }\sim 0$ contribute to tunneling if one
applies a bias. In our ASA calculations there are only surface resonant
states at $\epsilon _{F}$, and we have found localized states in the
minority channel about $k_{\parallel }\sim 0$ just below the Fermi level at 
$\epsilon \simeq \epsilon _{F}-0.05eV$. We also calculated the surface density
of states at $0.05eV$ below the Fermi level, as well as the
conductance $0.05eV$ below the Fermi level both at the interface and
in the bulk, i.e., $p=0$ and $4$ in terms of the in-plane momentum
$k_{\parallel }$. The large DOS at the surface about $k_{\parallel
}=0$ at $0.05eV$ below the Fermi level, which is absent at $\epsilon
_{F}$ and in the bulk $p=4$ indicates the presence of the localized
surface state. On comparing the conductances in, only the conductance
for the barrier $p=0$ and at $\epsilon _{F}-0.05eV$ has a strong
contribution from the localized states about $k_{\parallel }=0$; all
the other conductances have ``holes'' about $ k_{\parallel }=0$. One
notes that the average of the conductance for $p=0$ at $\epsilon
_{F}-0.05eV$ is four times larger than at $\epsilon _{F}$. The
conclusion that can be drawn is that if localized states exist about
$k_{\parallel }\sim 0$ for $E\sim \epsilon _{F}$ they would contribute
to the conductance measured across the barrier, but do not contribute
to the {\it ballistic} conductance away from the interfaces. In real
junctions where localized states are mixed with resonant and itinerant
at the surface they contribute to the conduction as measured across
electrodes far from the barrier; therefore it is only the ballistic
conductance calculated for the barrier itself, $p=0$, that captures
the contribution from localized states if they exist. In general the
contribution of localized states to tunneling will depend on their
coupling to the states in the barrier, i.e.\ their chemical bonding.

In conclusion, when one compares the conductance of real magnetic tunnel
junctions with diffusive electrodes to a calculation of conductance where it
is assumed that transport is ballistic throughout, e.g., for large $p$, we
find it overestimates the role of specular reflections as seen by the
electrodes, and, if they are present, overlooks contributions from states
localized near the interfaces that are coupled to itinerant states in the
electrodes by diffusive and relaxation processes in real planar junctions.
Both mechanisms present in all planar tunnel junctions conspire to maintain
the conductance relatively constant as we go away from the barrier, i.e.,
the decrease in the ballistic conductance does not apply to realistic planar
junctions. For these reasons the only conductance one obtains in a ballistic
calculation that can be compared to real junctions is that across the
barrier $p=0$. This should not suggest that ballistic calculations far away
from the barrier are meaningless; one can certainly think of systems such as
two iron whiskers separated by a vacuum or MgO~\cite{heinrich} barrier where
conductance in the electrodes is purely ballistic at very low temperatures.

We have used vacuum whereas the barriers in the tunnel junctions studied to
date have been insulators; while this changes the conductance one
calculates, it does not alter the conclusion we arrive at, i.e., if one does
a ballistic calculation it should be that of only the barrier. For finite
bias one probes a larger region about the Fermi level so that localized
surface states away from $\epsilon _{F}$ contribute to conduction; their
contribution to the conductance will be included if one does the calculation
directly across the barrier rather than in the electrodes. 

We would like to acknowledge and thank William Butler for sharing with us
his unpublished results on the tunneling conductance of Fe/vac/Fe junctions,
Matthias Bode for bringing to our attention his spin polarized tunneling
results, and Phivos Mavropolous and Nickos Papanikolaou for helpful
discussions. This work was supported by the Defense Advanced Research
Projects Agency and Office of Naval Research (Grant No. N00014-96-1-1207 and
Contract Nos. MDA972-96-C-0014, and MDA972-99-C-0009\ ), the National
Science Foundation (Grant No. INT-9602192), NATO\ (Grant No. CRG 960340),
the Hungarian National Science Foundation (OTKA T030240), and the TMR
Network ERBFMXCT-960089.

\end{document}